\documentclass[review]{elsarticle}

\usepackage{hyperref}  

\usepackage[mathlines]{lineno} 

\usepackage{placeins}
\usepackage{float}
\journal{Mechanical Systems and Signal Processing}

\usepackage{color}
\usepackage{amsmath}
\usepackage{amsfonts}
\usepackage{siunitx}

\graphicspath{{./Figs/}}
\begin{document}

\begin{frontmatter}

\title{Parameter reduction in nonlinear state-space identification of hysteresis}

\author[*]{Alireza Fakhrizadeh Esfahani\corref{mycorrespondingauthor}}
\cortext[mycorrespondingauthor]{Corresponding author: Alireza Fakhrizadeh Esfahani, Vrije Universiteit Brussel, department ELEC, 
Pleinlaan 2,
Building K, 6th floor,
B-1050 Brussels, Belgium } \ead{alireza.fakhrizadeh.esfahani@vub.ac.be}
\author{Philippe Dreesen$^1$}
\author{Koen Tiels$^1$}
\author{\\ Jean-Philippe No\"el$^{1,2}$}
\author{Johan Schoukens$^1$}
\address{$^1$%
Vrije Universiteit Brussel (VUB), 
Department ELEC, 
Brussels, Belgium}
\address{$^2$%
University of Li\`ege, Space Structures and Systems Laboratory,
Aerospace and Mechanical Engineering Department,
Li\`ege, Belgium}

\begin{abstract}\linenumbers
Hysteresis is a highly nonlinear phenomenon, showing up in a wide variety of science and engineering problems. 
The identification of hysteretic systems from input-output data is a challenging task. 
Recent work on black-box polynomial nonlinear state-space modeling for hysteresis identification has provided promising results, but struggles with a large number of parameters due to the use of multivariate polynomials. 
This drawback is tackled in the current paper by applying a decoupling approach that results in a more parsimonious representation involving univariate polynomials. 
This work is carried out numerically on input-output data generated by a Bouc-Wen hysteretic model and follows up on earlier work of the authors.
The current article discusses the polynomial decoupling approach and explores the selection of the number of univariate polynomials with the polynomial degree, as well as the connections with neural network modeling. 
We have found that the presented decoupling approach is able to reduce the number of parameters of the full nonlinear model up to about 50\%, while maintaining a comparable output error level. 
\end{abstract}

\begin{keyword}
Polynomial Nonlinear State-Space\sep Hysteretic System\sep Bouc-Wen\sep Tensor
Decomposition\sep Canonical Polyadic Decomposition\sep Decoupling Multivariate Polynomials
\end{keyword}

\end{frontmatter}


\section{Introduction}
Hysteresis is encountered in a wide range of engineering and scientific areas, and its analysis, identification and control are receiving increasing research interest in recent years \cite{bern,bern1,hassani2014survey,Ikhouane11,Noel,worden2007identification,worden2012identificationI,worden2012identificationII,worden2012parameterIII}. 
Hysteretic systems exhibit a nonlinear memory phenomenon, causing that effects of the input to the output are delayed in time.
Different from a phase delay, which is present in all linear dynamical systems but tends to zero for slowly varying input signals, hysteresis is characterized by an input-output looping behavior that persists when the forcing frequency approaches zero \cite{bern1}. 

Identification of hysteretic systems is important for estimation, modeling and control purposes \cite{Ikhouane11}, but poses a challenge because dynamic nonlinearities are governed by non-measurable internal state variables \cite{Ikhouane11,Noel}. 
Existing methods for identification of hysteretic systems have been considering mostly white-box approaches, and to a lesser extent black-box modeling principles. 
White-box modeling principles assume a hysteresis model, and need to be tailored towards a pre-specified model structure \citep{Ikhouane11,li2004improvement,worden2012parameterIII}, but may require prohibitively complicated models for real-life identification tasks. 
Black-box identification methods \citep{billings2013nonlinear,schon2011system,Johan1,Johan2,nelles2013nonlinear,Paduart1} have recently been proposed for the identification of hysteretic systems \cite{Noel,worden2007identification}, and perform identification by finding some model that provides an accurate description of the input-output data, such as neural networks \cite{Anna,xie2013identification} or polynomial state-space models \cite{Paduart1,Noel}. 
Black-box procedures provide powerful models, but they lack to provide physical intuition, and they often require a (very) large number of parameters. 
Therefore any method which can decrease the number of parameters is of interest.

We develop an approach that aims at combining the benefits of white-box and black-box approaches. 
We start by identifying a polynomial nonlinear state-space (PNLSS) model for the hysteretic system. 
We use the method of \cite{Noel}, which starts by extracting the best linear approximation \cite{Johan1} of the system and estimates the nonlinear distortion level of the system response. 
From this starting point, a polynomial state-space model \cite{Paduart1} is identified, which is able to reduce the output error to near the noise level, but requires a large number of parameters \cite{Noel}. 
We then apply a decoupling procedure that expresses the nonlinear part of the state-space model using univariate (one-to-one) polynomials in linear combinations of states and inputs. 
The decoupling method that is used here is adopted from \citep{Philippe111} and uses the canonical polyadic decomposition (CPD) of a three-way tensor \citep{Carrol,Harshman,Kolda}.
This decoupled representation is able to reduce the number of parameters significantly, and results in an interpretable model \cite{Fakhrizadeh}.
For the identification of the model we use input and output data generated by a Bouc-Wen hysteresis model \cite{Bouc,wen1976method,Noel}, which were provided in the recent nonlinear system identification benchmark challenge \cite{Maarten}.
This work follows up on the article \citep{Fakhrizadeh} by the same authors, in which we used a tensor decomposition approach to find a decoupled polynomial nonlinear state-space model with fewer parameters, and which resulted in a model with a slightly higher rms error than the full PNLSS model. 
In the current article, an algorithm is presented that manages to find decoupled models having a better rms errors than the full PNLSS model. 
Also an additional approach is presented to start from a lower degree decoupled model and reach a higher degree decoupled model. This paper solely focuses on parameter reduction and not on the interpretation of the decoupled models.

The paper is organized as follows. 
In Section \ref{Sec:NLSS} the PNLSS structure and the identification method are introduced. 
Section \ref{Sec:Dec} discusses the decoupling procedure. 
In Section \ref{Sec:Res} the results are presented and a variation of the PNLSS model is used to get a better model faster and more efficiently. 
Section \ref{Sec:Concl} is devoted to conclusions and future work.

\section{Nonlinear state-space modeling}\label{Sec:NLSS}
In this section, nonlinear state-space modeling is introduced. The polynomial nonlinear state-space (PNLSS) model structure is introduced in Section \ref{PNLSS}. Its connection with the Bouc-Wen hysteretic model is elaborated on in Section \ref{Sec:Hyster}. The PNLSS identification algorithm is explained briefly in Section \ref{PNLSSID}.

\subsection{Polynomial nonlinear state-space (PNLSS) models\label{PNLSS}}
Let us start by motivating the use of discrete-time polynomial nonlinear state-space models. 
Discrete-time polynomial nonlinear state-space models are introduced in \citep{Paduart1,Paduart3,Paduart4}.
The estimation of these discrete-time models is computationally less involved than that of their continuous-time counterparts, and for control applications, discrete-time models are more suitable \citep{Paduart4}. This motivates the choice for discrete-time models.
Polynomials are straightforward to use in computations and it is easy to extend them for multivariable applications \citep{Paduart1}, which motivates the choice for polynomials.

In this paper we only consider single input single output (SISO) systems. For a SISO system the polynomial nonlinear state-space model is
\begin{eqnarray}
x(t+1) &=& A x(t)  + b u(t) + E \ \zeta (x(t),u(t)), \nonumber \\
y(t)&=& \! \!  c^T x(t) + d u(t)+f^T \eta (x(t),u(t)) \label{eq:FullPNLSS}
\end{eqnarray}
where $u(t) \in \mathbb{R}$, $y(t) \in \mathbb{R}$ and $x(t) \in \mathbb{R}^n$ are the input, output and state vector at time instance $t$, respectively. The linear part of the state-space model is composed of $A \in \mathbb{R}^{n \times n}$ (the state transition matrix), $b \in \mathbb{R}^{n}$, $c \in \mathbb{R}^{n}$, and $d \in \mathbb{R}$.
The vectors $\zeta(x(t),u(t)) \in \mathbb{R}^{n_\zeta}$ and $\eta(x(t),u(t)) \in \mathbb{R}^{n_\eta}$ contain all possible monomials in the states and input of degrees two up to $d$ (the linear terms are already captured by the linear state-space part).
The matrix $E \in \mathbb{R}^{n \times n_{\zeta}}$ and $f \in \mathbb{R}^{n_{\eta}}$ contain the corresponding polynomial coefficients.
As an illustration, for a second-order system with one input the monomials of
degree two are
\begin{equation}
 \zeta(x,u) =
 \eta(x,u)=
\begin{bmatrix} x_1^2 & x_1 x_2 & x_1 u & x_2^2 & x_2 u & u^2 \end{bmatrix}^T. 
\end{equation}
It can be shown that the total number of nonlinear monomials is given by \citep{Paduart1}
\begin{equation}
\left(
\frac{(n+1+d)!}{d!(n+1)!}-n-2
\right)
(n+1).
\label{eq:Comb}
\end{equation}

\subsection{PNLSS identification algorithm\label{PNLSSID}}
The polynomial nonlinear state-space identification algorithm is introduced in \citep{Paduart1} and further developed by \citep{Laurent1, Laurent2,Paduart3,Anna}. The algorithm is explained in details in \citep{Paduart4,Paduart1}. Here the procedure is introduced briefly.
\begin{itemize}
\item[1.] Calculate the best linear approximation (BLA)  $\hat{G}_{BLA}$ from the input-output data and its total covariance  $\hat{\sigma}_{G_{BLA}}$ using robust method \citep{Johan1,Johan2}.
\item[2.] From $\hat{G}_{BLA}$, $\hat{\sigma}_{G_{BLA}}$, and by using a linear subspace identification \citep{Rik1,van2012subspace} method, find an initial estimate for $A$, $b$, $c$, and $d$.
\item[3.] The parameters $A$, $b$, $c$, $d$, $E$, and $f$ are calculated by Levenberg-Marquardt optimization algorithm by minimizing the (weighted) mean-square output error.
\end{itemize}

\subsection{The Bouc-Wen model for hysteretic systems}\label{Sec:Hyster}
The model which is used here to simulate a hysteretic system is a Bouc-Wen model \citep{kyprianou2001identification,wen1976method}. The differential equations of this model are
\begin{eqnarray}
&&m_L \ddot{y} + c_L \dot{y} + k_L y + z(y,\dot{y}) = u(t) \nonumber \\
&&\dot{z}(y,\dot{y}) = \alpha \dot{y} - \beta (\gamma \left | \dot{y} \right | \left | z \right | ^{\nu -1}z + \delta \dot{y} \left | z \right |^{\nu})\label{eq:BW}
\end{eqnarray}
where $m_L$, $k_L$, and $c_L$ are the linear mass, stiffness and viscous damping coefficients, respectively. The coefficients $\alpha$, $\beta$, $\gamma$, $\delta$, and $\nu$ are the Bouc-Wen parameters, which are used to tune the shape and the smoothness of the hysteresis loop.
The Bouc-Wen model parameters are listed in Table \ref{tab:PARS}. 
Newmark integration \citep{newmark1959method,geradin2014mechanical} is used to simulate the behavior of the Bouc-Wen model \eqref{eq:BW} and is explained in details in \citep{Noel,Maarten} for this specific problem.

The Bouc-Wen model in \eqref{eq:BW} can be represented in a nonlinear state-space form \citep{Noel}
\begin{equation}
\begin{aligned}
\begin{bmatrix}
\dot{y} \\ \ddot{y} \\ \dot{z}
\end{bmatrix}
& =
\begin{bmatrix}
0 & 1 & 0 \\
-\frac{k_L}{m_L} & -\frac{c_L}{m_L} & -\frac{1}{m_L} \\
0 & \alpha & 0
\end{bmatrix}
\begin{bmatrix}
y \\ \dot{y} \\ z
\end{bmatrix}
+
\begin{bmatrix}
0 \\ \frac{1}{m_L} \\ 0
\end{bmatrix}
u +
\begin{bmatrix}
0 & 0 \\
0 & 0 \\
-\beta \gamma & -\beta \delta
\end{bmatrix}
\begin{bmatrix}
|\dot{y}|z \\ \dot{y}|z|
\end{bmatrix}\\
y & =
\begin{bmatrix}
1 & 0 & 0
\end{bmatrix}
\begin{bmatrix}
y \\ \dot{y} \\ z
\end{bmatrix}
\end{aligned}
\end{equation}
Although these are continuous-time state-space equations with non-polynomial nonlinearities, a discrete-time polynomial nonlinear state-space model can still obtain promising results \citep{Noel}.
\begin{center}
\begin{table}
\caption{Parameters of the Bouc-Wen model for data generation.}\label{tab:PARS}
\begin{tabular}{c|cccccccc}
\hline
Parameter & $m_L$ & $c_L$ & $k_L$ & $\alpha$ & $\beta$ & $\gamma$ & $\delta$ & $\nu$ \\
Values (in SI unit) & \SI{2} & \SI{10} & \SI{5e4} & \SI{5e4}&\SI{1e3} & \SI{0.8}& \SI{-1.1} & \num{1} \\
\hline
\end{tabular}
\end{table}
\end{center}
\subsection{Earlier obtained results}
Since the current paper relies a lot on \citep{Noel}, the results of that paper are briefly reviewed here. In \citep{Noel}, the signal used for training the model has an rms value of $50$~N. The other characteristics (sampling frequency, frequency lines of excitation and the number of samples) are the same as in this work. The PNLSS model contains all possible monomials of the states and the input up to the chosen polynomial degree. The results are shown in Table \ref{tab:MSSP}. It can be seen that the number of parameters is growing in a combinatorial manner. The PNLSS model with a polynomial degree of $3-5-7$ has the minimum error. For this work the PNLSS model of polynomial degree of $2-3$ is estimated from a multisine of $55$~N.

\begin{center}
\begin{table}
\caption{RMS error on validation data for polynomial nonlinear state-space models of various degrees together with their respective number of parameters. Results from \citep{Noel}.}
\begin{tabular}{lcc}
\hline
Polynomial degree & RMS validation error (in dB) & Number of parameters\\
$2$ & $-85.32$ & 34 \\
$2-3$ & $-90.35$ & 64 \\
$2-3-4$ & $-90.03$ & 109 \\
$2-3-4-5$ & $-94.87$ & 172 \\
$2-3-4-5-6$ & $-94.85$ & 256 \\
$2-3-4-5-6-7$ & $-97.96$ & 364 \\
$3-5-7$ & $-98.32$ & 217 \\
\hline
\end{tabular}\label{tab:MSSP}
\end{table}
\end{center}

\section{Parameter reduction by decoupling}\label{Sec:Dec}
\subsection{Decoupled nonlinear state-space\label{DECPNLSS}}
The number of parameters in the full polynomial nonlinear state-space model given by \eqref{eq:Comb} can become very large for large model orders $n$ and/or large nonlinear degrees $d$.

\begin{figure}
\begin{center}
\includegraphics[scale = .29, trim = 1.3cm 1.5cm 0 1.5cm]{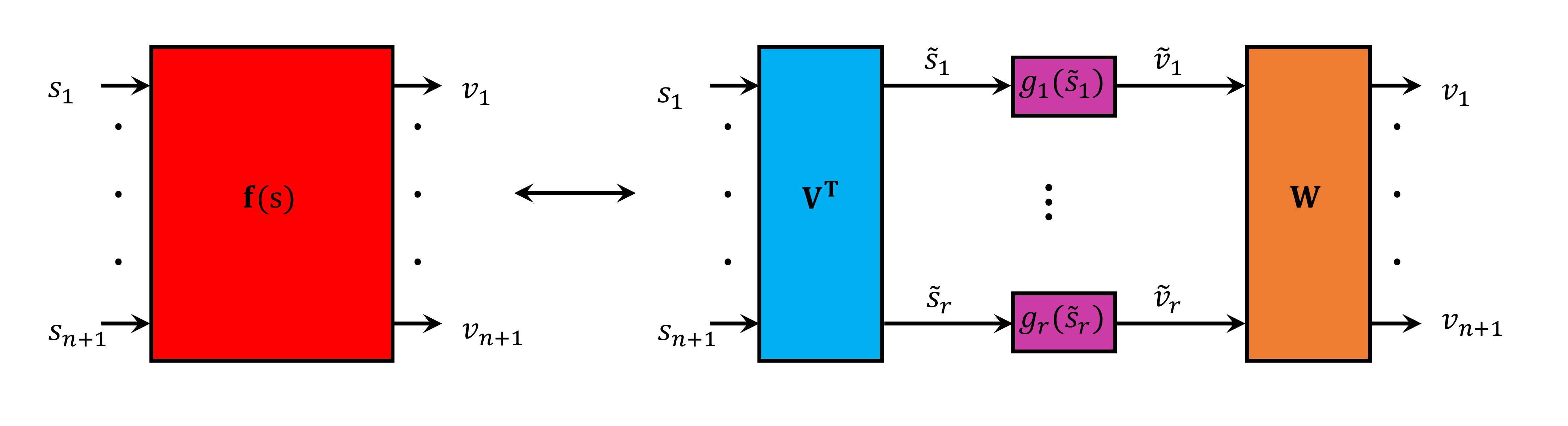}    
\caption{The concept of decoupling the model. A multivariate polynomial vector function is decomposed into a linear transformation $V$, followed by a set of parallel univariate polynomials $g_1,\ldots,g_r$, and another linear transformation $W$.} 
\label{fig:CPD}
\end{center}
\end{figure}

The polynomial nonlinear state-space can be simplified significantly by rotating the states and inputs. 
Indeed, often it is possible to find a new set of linearly transformed states and inputs in which the nonlinear part can be expressed in terms of only a few univariate polynomials. 
The decoupled polynomial nonlinear state-space model is as follows
\begin{eqnarray}
x(t+1) &=& Ax(t) + b u(t) + W_x g \begin{pmatrix}V^T \begin{bmatrix} x(t) \\ u(t) \end{bmatrix} \end{pmatrix}, \nonumber \\
y(t) &=& \! \! c^T x(t) + d u(t) +  w_y^T g \begin{pmatrix} V^T \begin{bmatrix} x(t) \\ u(t) \end{bmatrix} \end{pmatrix},
\end{eqnarray}
where $W_x$ and $w_y^T$ represent the linear transformations  for transforming the nonlinear univariate polynomial functions in the PNLSS equation and $V$ is the transformation matrix for transforming states and inputs (see Figure~\ref{fig:CPD}.
The system of multivariate polynomials ${f}(s)$ with inputs $s_1, s_2, \cdots s_{n+1}$ is transformed into the new representation which is composed of a linear transformation matrix ${V^T}$, $r$ branches of univariate polynomials $g_1(\tilde{s}_1), \cdots , g_r(\tilde{s}_r)$, and a linear transformation matrix ${W}$ to transform back to the original output space.

\subsection{Decoupled nonlinear state-space model identification algorithm}
\label{sec:DecouplingAlgorithm}
An initial decoupled model can be found from the canonical polyadic decomposition of the Jacobian of the multivariate functions in the PNLSS model. 
The parameters of this initial decoupled model then can be further optimized using Levenberg-Marquardt optimization. The following steps present the detailed algorithm:
\begin{itemize}
\item[1.] The Jacobian of the polynomial vector function $f(s)$ is evaluated in a set of $N$ sampling points $s^{(k)}$, for $k = 1,2, \ldots N$, which are drawn from a random normal distribution. 
\item[2.] The Jacobian matrices are stacked into an $(n+1) \times (n+1) \times N$ tensor (see Fig.~\ref{fig:StackingGrads}). 
We have thus 
\begin{equation}
J_{ijk}=\frac{\partial v_i (s_j^{(k)})}{\partial s_j}.
\end{equation}
\item[3.] Estimate the rank $r$ of tensor $J_{ijk}$. 
This is done by scanning a number of candidate values for $r$ and selecting the one for which the approximation error of the CPD is sufficiently small. 
The rank $r$ corresponds to the number of branches. 
In the exact case, assuming that an underlying decomposition exists and is sufficiently generic (see \cite{Philippe111}), the decoupling task has a unique solution if 
\begin{equation}
n^2(n^2-1) \geq 2r(r-1).
\end{equation}
\item[4.] The tensor $J$ is decoupled using the canonical polyadic decomposition CPD as follows (see the figure below) \\
\begin{figure}[ht]
\begin{center}
\includegraphics[trim = 0cm 4cm 1cm 0cm, scale = .34 ]{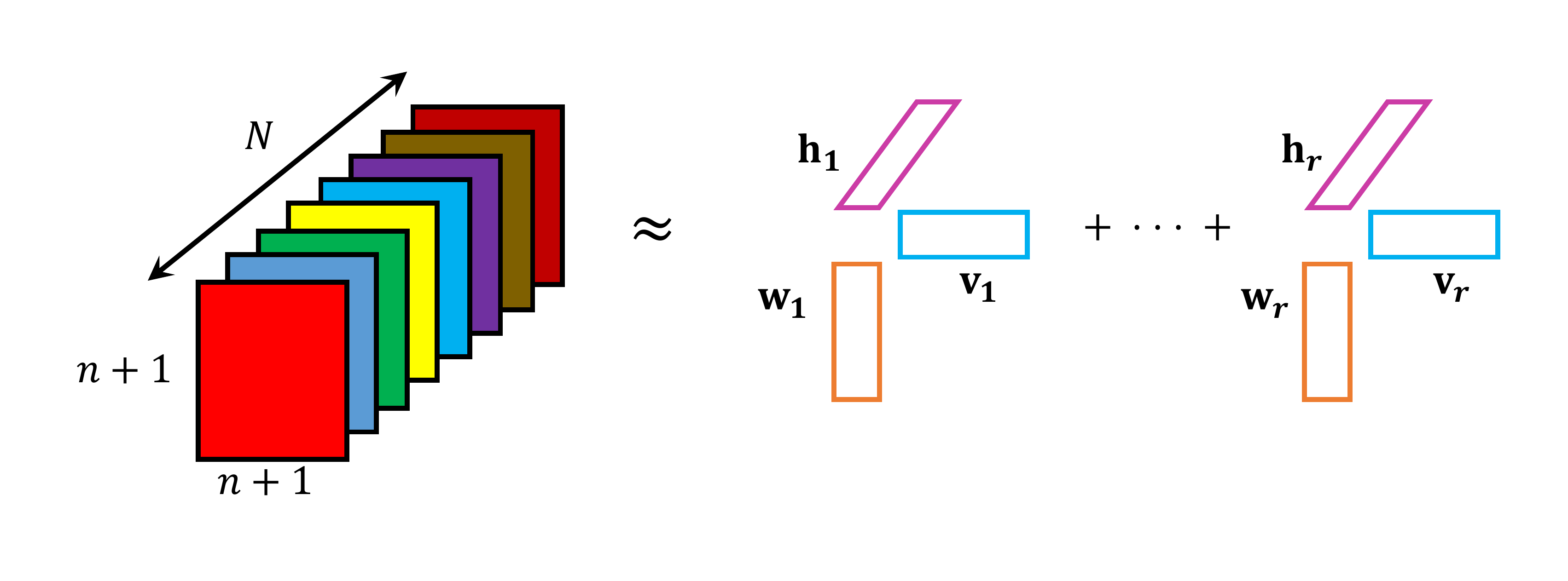}    
\label{fig:StackingGrads}
\end{center}
\end{figure}
\begin{equation}
J_{ijk} \approx \sum_{\ell = 1}^{r} \, w_{i\ell} \, v_{j\ell} \, h_{k\ell},\label{eq:DECEQ}
\end{equation}
where $W$ and $V$ are the CPD factors and
$h_{k\ell}= {g_\ell}'(\tilde{s}_\ell^{(k)})%
$
is the derivative of the univariate function $g_\ell$ evaluated in sampling points $\tilde{s}_\ell^{(k)}$, for $k = 1, \ldots, N$. 
\item[5.] For the $\ell^{th}$ branch ${g}'_\ell(\tilde{s}_\ell)$, we solve the following polynomial fitting
\begin{equation}
\begin{bmatrix}
(\tilde{s}_{\ell}^{(1)})^1 & (\tilde{s}_{\ell}^{(1)})^2 & \cdots & (\tilde{s}_{\ell}^{(1)})^{d-1} \\
(\tilde{s}_{\ell}^{(2)})^1 & (\tilde{s}_{\ell}^{(2)})^2 & \cdots & (\tilde{s}_{\ell}^{(2)})^{d-1} \\
\vdots && \vdots\\
(\tilde{s}_{\ell}^{(N)})^1 & (\tilde{s}_{\ell}^{(N)})^2 & \cdots & (\tilde{s}_{\ell}^{(N)})^{d-1}
\end{bmatrix}
\begin{bmatrix}
{c}'_{\ell,1} \\
{c}'_{\ell,2} \\
\vdots \\
{c}'_{\ell,d-1}
\end{bmatrix}= 
\begin{bmatrix}
{h}_{1\ell} \\
{h}_{2\ell} \\
\vdots \\
{h}_{N\ell} \\
\end{bmatrix},\label{Eq:DECREGR}
\end{equation}
leading to the coefficients of $g'_\ell$. The constant and linear terms are not considered. 
The symbolic integration
\begin{equation}
g_\ell(\tilde{s}_\ell) = \int {g}'_\ell(\tilde{s}_\ell)d\tilde{s}_\ell,\label{eq:Univars}
\end{equation}
determines the functions $g_\ell$ up to the correct value of the integration constants.
\item[7.] This algorithm, till this point gives an initialization for the decoupled model. If this initial decoupled model encounters stability problems during further optimization when it is plugged into the PNLSS model structure, it is needed to initialize the optimization algorithm from a stable linear model. This initial model then still contains the univariate polynomial coefficients (coefficients of $g_{\ell}(\tilde{s}_{\ell})$ in \eqref{eq:Univars}) and the matrix $V$, but the matrix $W$ is set to zero to make the output of the nonlinear part in this initial decoupled model zero.
\item[8.] We use the Levenberg-Marquardt optimization algorithm to minimize the (weighted least square (WLS)) output error, initialized with this model. The decoupled model is iterated until the optimal solution is obtained.
\item[9.] Go to step 1 and regenerate another set of random points. Try several times and select the model which gives the minimum amount of the error on the validation data set among all candidates.
\end{itemize}

The obtained model has ($(n+1) r + (n+1) r + r  (d-1) = (2n+d+1) r$) nonlinear parameters terms. 
It is worth mentioning two benefits of this approach:
\begin{itemize}
\item The number of parameters increases linearly with the degree of nonlinearity ($d$), which is in contrast to the combinatorial increase in the full polynomial state-space approach.
\item The reduction of number of estimated parameters allows for the ability to increase the degree of the nonlinearity in the estimated model.
\end{itemize}

\subsection{Comparison with neural network}
According to the theorem of Weierstrass \citep{jeffreys1956methods} every continuous function can be approximated arbitrarily well by polynomials on a closed and bounded interval. 
Parameters of nonlinear models expressed by polynomials can be optimized effectively. 
One drawback of using polynomials is the high orders for approximating some functions which is not the case in this study where the maximum order of polynomials is $11$. Another drawback is their poor extrapolation ability. 
The latter disadvantage of using polynomials is not the case for neural networks, however, neural networks have intrinsically the problem of local minima trap \citep{janczak2004identification}. In the full PNLSS model, the estimation of the polynomial coefficients is nonlinear in the parameters, and is hence also sensitive to local minima. In the estimation of the derivatives of the univariate functions $g_\ell$ \eqref{Eq:DECREGR}, the polynomial model can be expressed as one that is linear in the parameters. Although the tensor decomposition in \eqref{eq:DECEQ} is nonlinear in parameters, it helps that the problem is multilinear in the parameters.

The similarity between the PNLSS models (full and decoupled) and neural networks approaches is that they are typically black-box approaches. In \citep{xie2013identification}, however, the hysteresis Bouc-Wen model is discretized, and based on this model a neural network is designed to estimate the Bouc-Wen parameters. Because the underlying system equations are used, this approach seems a grey-box approach which is not comparable with the current black-box approaches.
\section{Results}\label{Sec:Res}
In this section the excitation and technical settings are defined, and results are explained and compared to some other studies that have been done.
\subsection{The excitation signal\label{Sec:Exc}}
For exciting the system a few points are taken into account.
\begin{itemize}
\item To be able to capture at least the $3^{rd}$ order nonlinearity, the highest excitation frequency  is selected to be at least three times as large as the resonance frequency (The resonance frequency can be estimated by a simple swept sine experiment).
\item The test duration is set by the frequency resolution.
\item To avoid extrapolation problems in the estimated model, the amplitude (rms level) of the input signal is slightly higher in the training set than in the test set.
\end{itemize}
For these reasons a random phase multisine \citep{Johan1} with the following property is chosen for the excitation. The standard deviation of the training multisine data is $55$~N and the standard deviation of the test data is $50$~N.
The random-phase multisine is generated through the following equation
\begin{equation}
u(t)=N_s^{-\frac{1}{2}}\sum_{k=-\frac{N_s}{2}+1}^{k=\frac{N_s}{2}-1}U_ke^{j(2\pi k \frac{f_s}{N_s}t+\phi_k)}
\end{equation}
where $U_k$ is the amplitude of each frequency line $k$. The phases $\phi_k$ are chosen randomly and uniformly from the interval $(-\pi \ \pi)$. Furthermore, $N_s$  is the number of time samples per period and is equal to $8192$, $f_s$ is the sampling frequency which is $750$~Hz, and $j$ is the imaginary unit.  Here the amplitudes $U_k$ in the excited frequency band are equal. By increasing the number of frequency lines the distribution of this signal approaches to a Gaussian distribution. The frequency band of excitation is $[5 \ - \ 150]$~Hz.

Moreover, a swept-sine test dataset is available with a sweep from $20$ to $50$~Hz at a rate of $10$~Hz per minute \citep{Maarten}.
\subsection{Decoupled models}
The system is excited with the signal which is generated with the properties mentioned in Section \ref{Sec:Exc}. The input/output data is collected. The PNLSS model is estimated with quadratic and cubic nonlinearities in the state equation. The Jacobian matrix of $E \ \zeta (x,u)$  is evaluated at $N = 500$ points. The rank of the Jacobian tensor $J_{ijk}$ is estimated by the $\texttt{rankest}$ command of the Tensorlab toolbox \citep{TensorLab}. It calculates the rank of the tensor which is defined as the minimum number of rank-one tensors that generate the tensor as their sum \citep{Kolda}. This rank is estimated to be $6$. It suggests to check all possible number of branches up to $6$. The maximum degree of univariate polynomials is checked from $2$ to $11$. All these possible models are re-estimated $5$ times with different random grid points for Jacobians. The generated models are tested on the test data.
\subsubsection{Overview of all decoupled models}
\begin{figure}
\begin{center}
\includegraphics[scale = .31, trim =1cm 1.5cm 0 1.5cm]{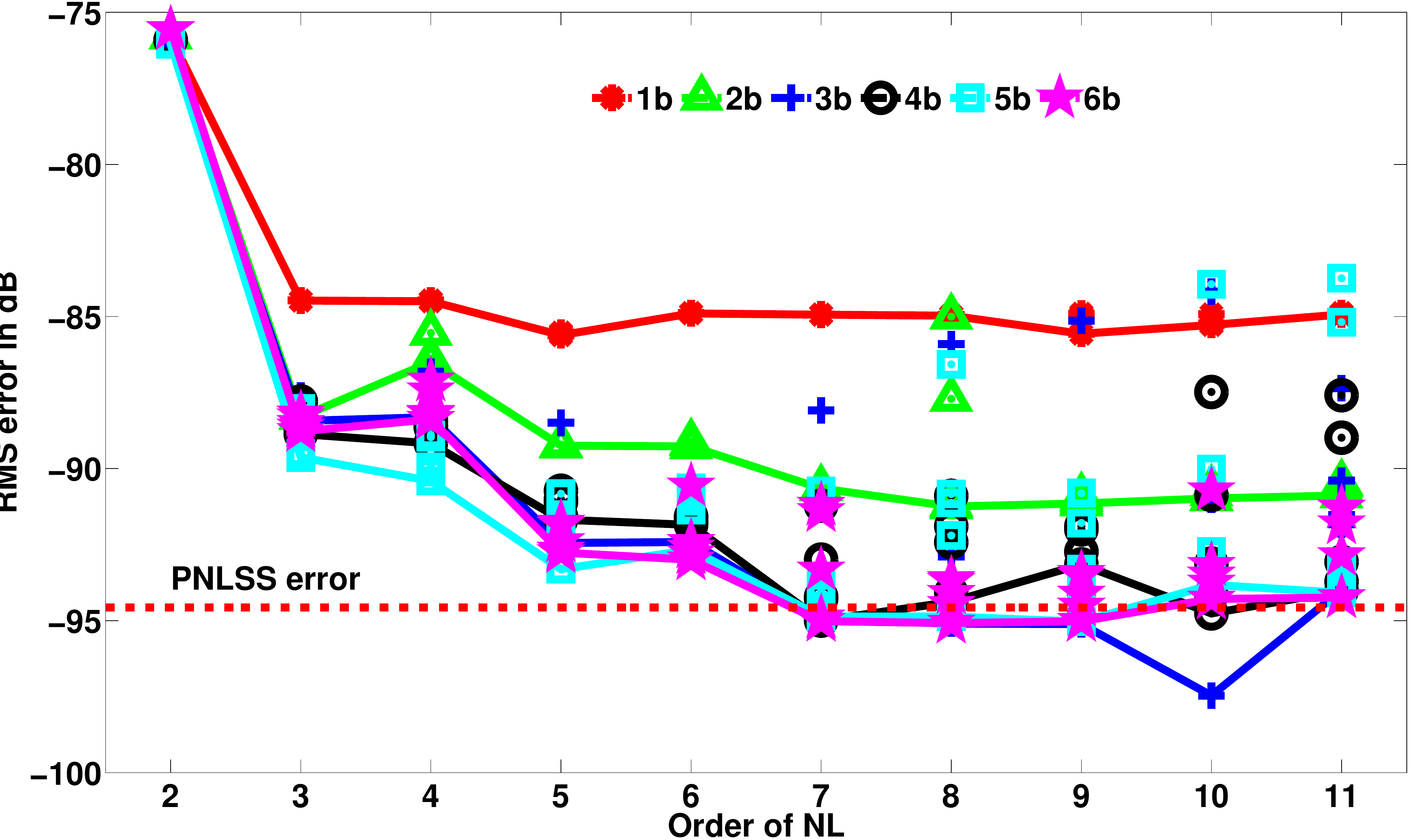}    
\caption{The rms error (in dB) on the multisine test data of all candidate models with different number of branches and different order of nonlinearity in univariate polynomial branches. $1b,\ 2b, \ \cdots \ 6b$ stand for $1, \ 2, \ \cdots \ 6$ branches.
The best decoupled model has 3 branches and $10^{th}$ order of nonlinearity, and achieves a significant error reduction compared to the full PNLSS model.}
\label{fig:RMS1}
\end{center}
\end{figure}

\begin{figure}[ht]
\begin{center}
\includegraphics[scale = .31, trim =1cm 1.5cm 0 1.5cm]{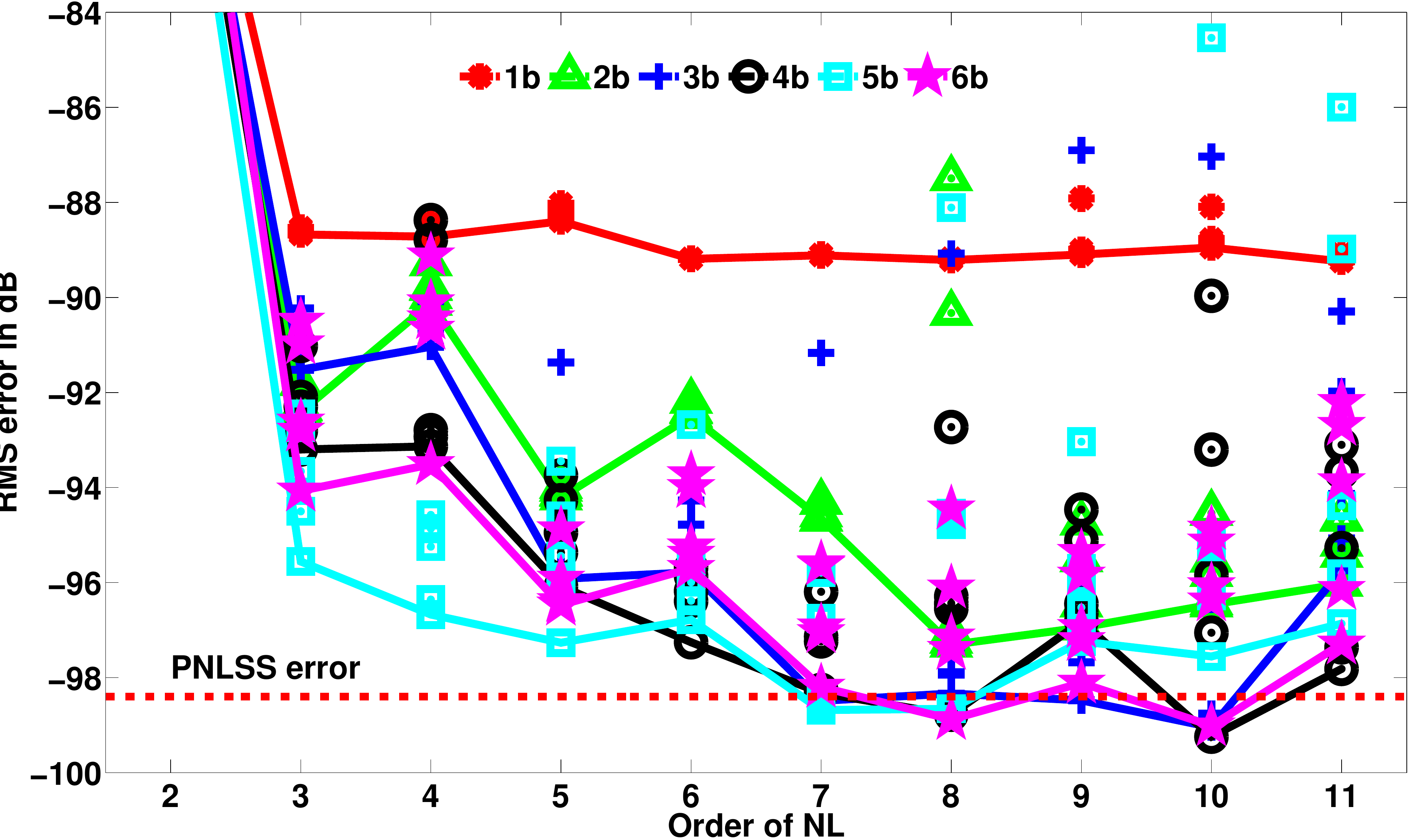}    
\caption{The rms error (in dB) on the swept--sine test data of all candidate models with different number of branches and different order of nonlinearity in univariate polynomial branches. $1b,\ 2b, \ \cdots \ 6b$ stand for $1, \ 2, \ \cdots \ 6$ branches.
Also for the swept-sine test data, to obtain an error less than or similar to the full PNLSS model, the decoupled model needs at least 3 branches and $7^{th}$ order of nonlinearity.} 
\label{fig:RMS2}
\end{center}
\end{figure}

Figure \ref{fig:RMS1} shows the rms error (in dB) of all generated decoupled models vs. the order of the decoupled univariate polynomials on the multisine test data. Figure \ref{fig:RMS2} shows a similar plot for the swept-sine test data. For each decoupled model all $5$ estimated models are depicted with a marker. The minimum of each group of five trials are connected with a line for models with the same number of branches. As a reference, the error of the full PNLSS model is shown with a dashed line. From Figure \ref{fig:RMS1} it can be observed that the rms error for all decoupled models with $2^{nd}$ order univariate polynomials is $20$~dB larger than that of the full PNLSS model. The one branch decoupled model converges to a fixed error level which is about $10$~dB higher than the PNLSS error. The decoupled model with 2 branches shows a slightly better error than the one branch model and converges to an rms error that is about $4$~dB larger than that of the full PNLSS model. Models with 3 branches up to 6 branches and $7^{th}$ to 11$^{th}$ order of polynomials show the error near the PNLSS error. Among these models the 3-branch model with $10^{th}$ order of nonlinearity with $51$ nonlinear terms gives an error significantly fewer than PNLSS error especially for the 
multisine (see Figure \ref{fig:RMS1}),
although this model shows a great variation from one trial to the other. This model is the best model, but largely depends on the CPD initialization which depends on the choice of points where Jacobian is calculated (see Section \ref{SubSec:TechIss}). It is worth to mention that this model is comparable to PNLSS models of nonlinear order of $2-3-4-5-6-7$ and $3-5-7$ (Table 3 in \citep{Noel}) where they are the best model of that work, but with $364$ and $217$ parameters, respectively. All the decoupled models with polynomials of order $11$ show higher error than PNLSS. It seems from this model order that the decoupled model starts to overfit.

\begin{figure}[ht]
\begin{center}
\includegraphics[scale = .31, trim =0.8cm 1.5cm 0 1.5cm]{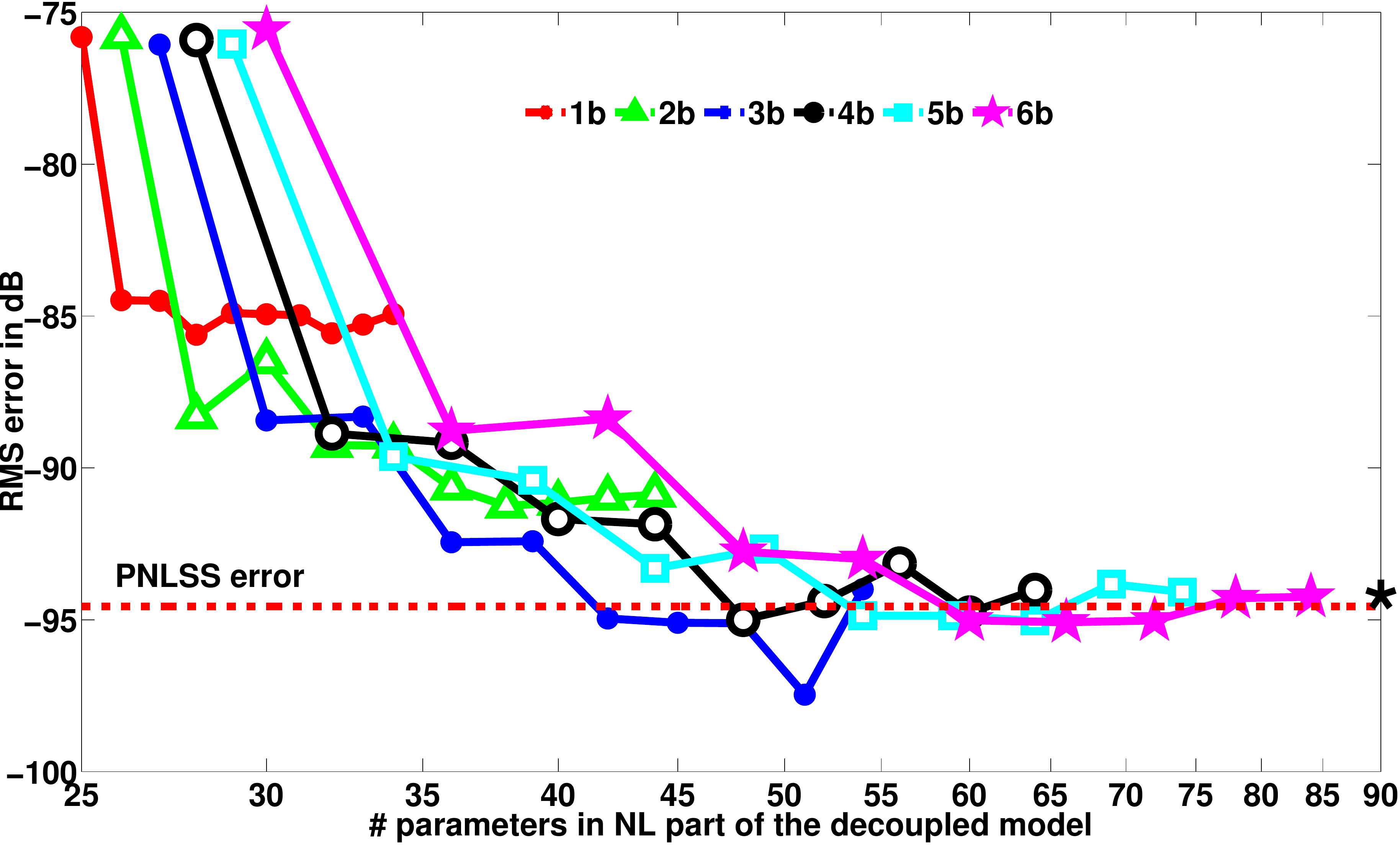}    
\caption{The minimum rms error (in dB) on the multisine test data of all candidate models with different number of branches and different number of parameters in the decoupled model (the scale of horizontal axis is logarithmic). $1b,\ 2b, \ \cdots \ 6b$ stand for $1, \ 2, \ \cdots \ 6$ branches. The black star in the far right shows the number of parameters of the  PNLSS model (90 parameters).
A significant parameter reduction can be achieved with a (sufficiently flexible) decoupled model while maintaining or even improving on the rms error of the full PNLSS model.}
\label{fig:RMS3}
\end{center}
\end{figure}

\begin{figure}[ht]
\begin{center}
\includegraphics[scale = .31, trim =.8cm 1.5cm 0 1.5cm]{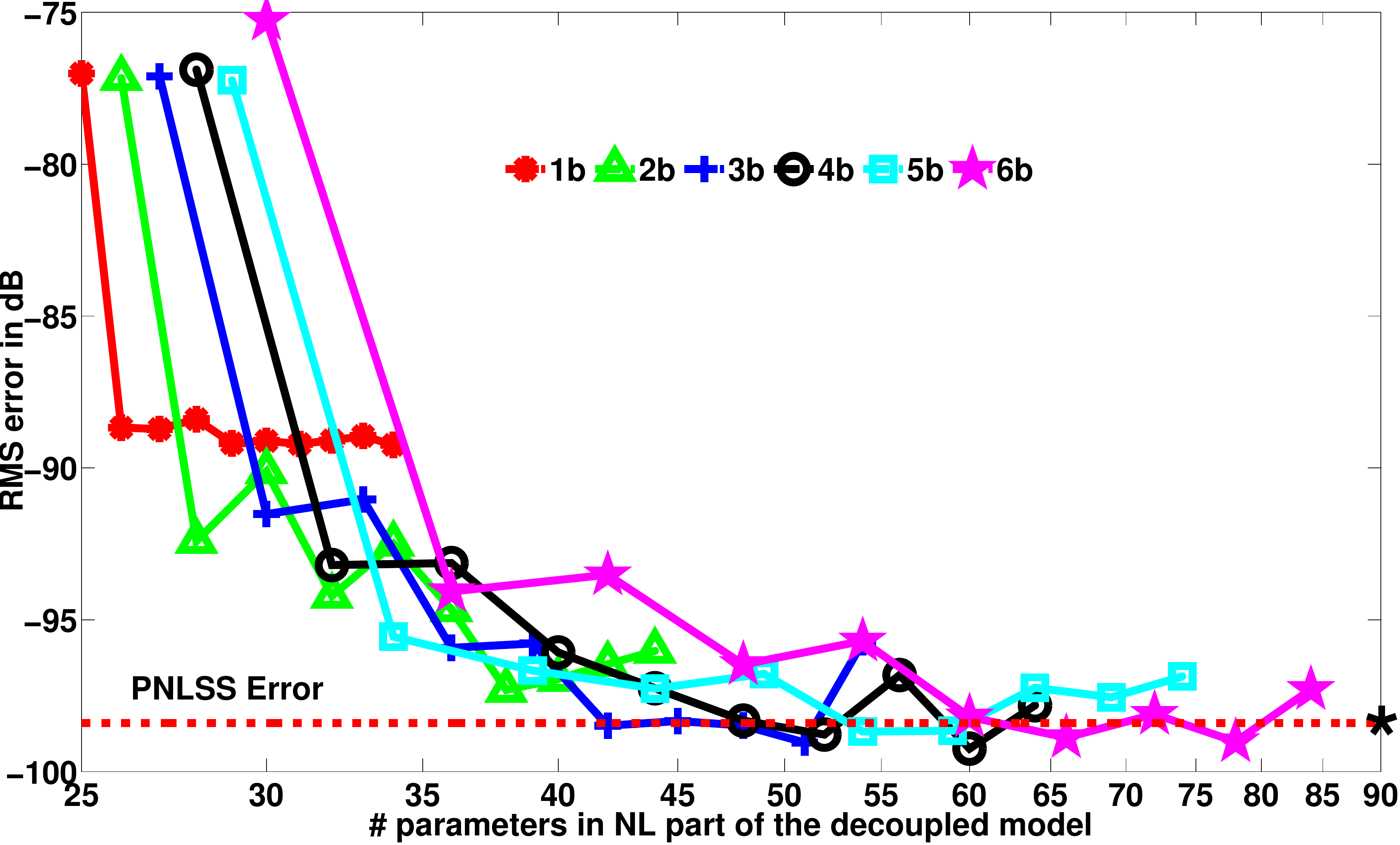}    
\caption{The minimum rms error (in dB) on the swept--sine test data of all candidate models with different number of branches and different number of parameters in the decoupled model (the scale of horizontal axis is logarithmic). $1b,\ 2b, \ \cdots \ 6b$ stand for $1, \ 2, \ \cdots \ 6$ branches. The black star in the far right shows the number of parameters of the  PNLSS model (90 parameters).
A significant parameter reduction can be achieved with a (sufficiently flexible) decoupled model while maintaining or even improving on the rms error of the full PNLSS model.} 
\label{fig:RMS4}
\end{center}
\end{figure}

The same information is shown in Figures \ref{fig:RMS3} and \ref{fig:RMS4}, but in a different way. The horizontal axis shows the number of nonlinear terms in a logarithmic scale. The decreasing trend in the error of decoupled model is quite obvious.
In Figure \ref{fig:RMS3}, models with less than 42 parameters have a higher error than the full PNLSS model. From 42 parameters onwards (3-branch model with $7^{th}$ order nonlinearities), we see that the decoupling approach reduces the number of parameters significantly (the full PNLSS has 90 nonlinear parameters), while most decoupled models still have enough flexibility to capture the behavior of the system. From 54 parameters onwards, the decoupled models are more flexible and they all have an error comparable to that of the full PNLSS model.
From this figure, all $2^{nd}$ order models have nearly the same error. The rms error for the estimated linear model is $-76$~dB which means the $2^{nd}$ order model has the same error level as the linear model. This fact reminds that the system has a significant odd nonlinearity \citep{Noel}. The systems with $2$ branches and with nonlinear order higher than $2$ are stagnated at $-91$~dB, without any significant improvement. This happens also for 1-branch models with nonlinear order higher than $2$. They stagnated around $-85$~dB. It can be seen that some models have the same number of parameters and the same error level, but a different number of branches and order of nonlinearity. For example, a 6-branch model with $3^{rd}$ order nonlinearity has the same number of parameters and error level as a 4-branch model with $4^{th}$ order nonlinearity. Similarly, a 5-branch model with $3^{rd}$ order nonlinearity and a 2-branch model with $6^{th}$ order nonlinearity have the same number of parameters and error level.

It is worth mentioning that the user should make a trade-off between precision and complexity. From Figure \ref{fig:RMS4}, it is seen the user has plenty of models which can be selected based on the rms error and complexity of the model.

\subsubsection{Result of the selected model}
\begin{figure}[ht]
\begin{center}
\includegraphics[scale = .34, trim =1cm 3cm 0 1.5cm]{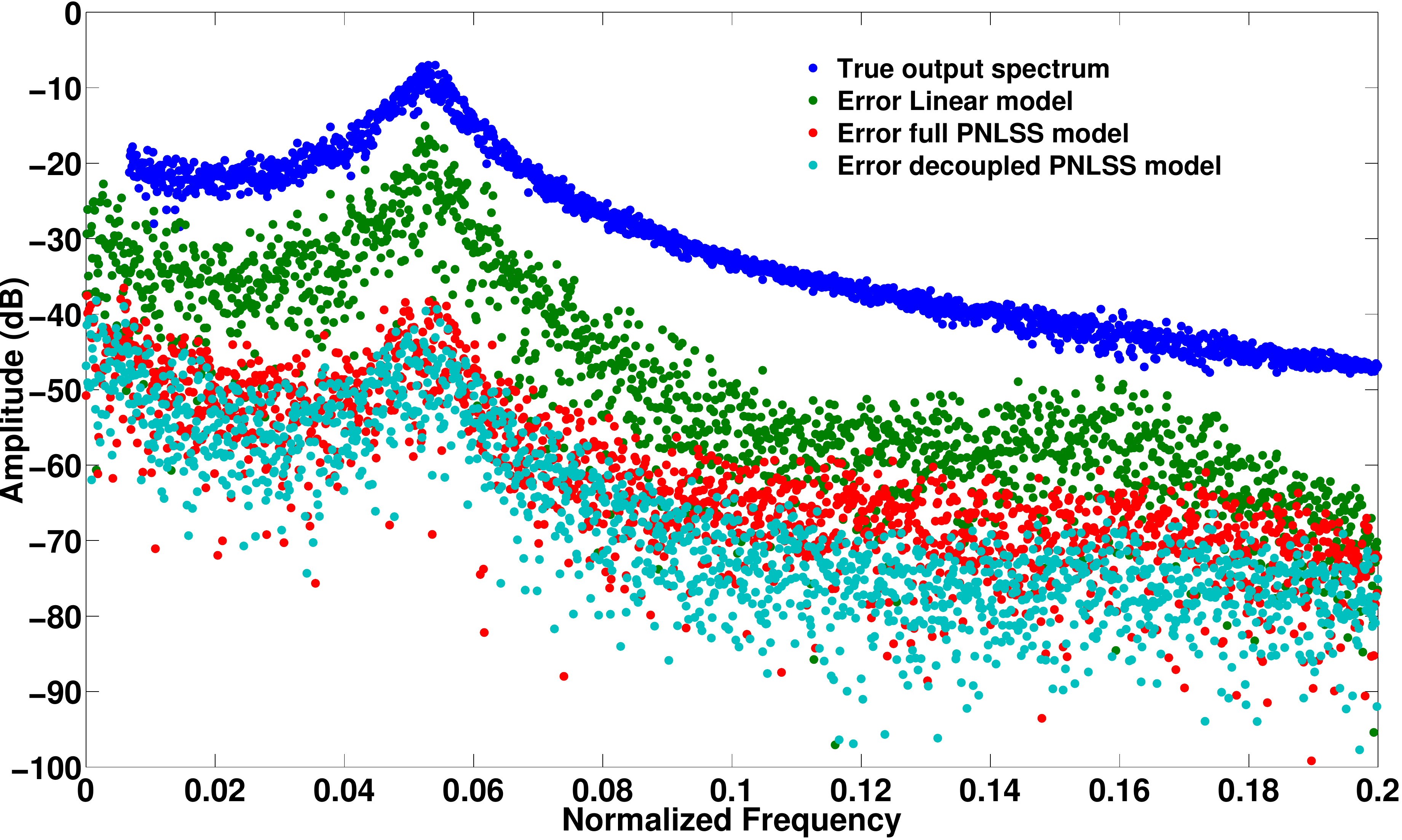}    
\caption{The spectrum of output error on the multisine test data of the selected decoupled model (in cyan) with 3 branches and nonlinear degree up to 10, the polynomial nonlinear state-space (PNLSS) model's output error (in red), the linear model's output error (in green), and true output (in blue).
The best decoupled model achieves its better performance compared to the full PNLSS model mainly at the higher frequencies.}
\label{fig:FRF}
\end{center}
\end{figure}

\begin{figure}[ht]
\begin{center}
\includegraphics[scale = .50, trim =1.3cm 3.3cm 0 1cm]{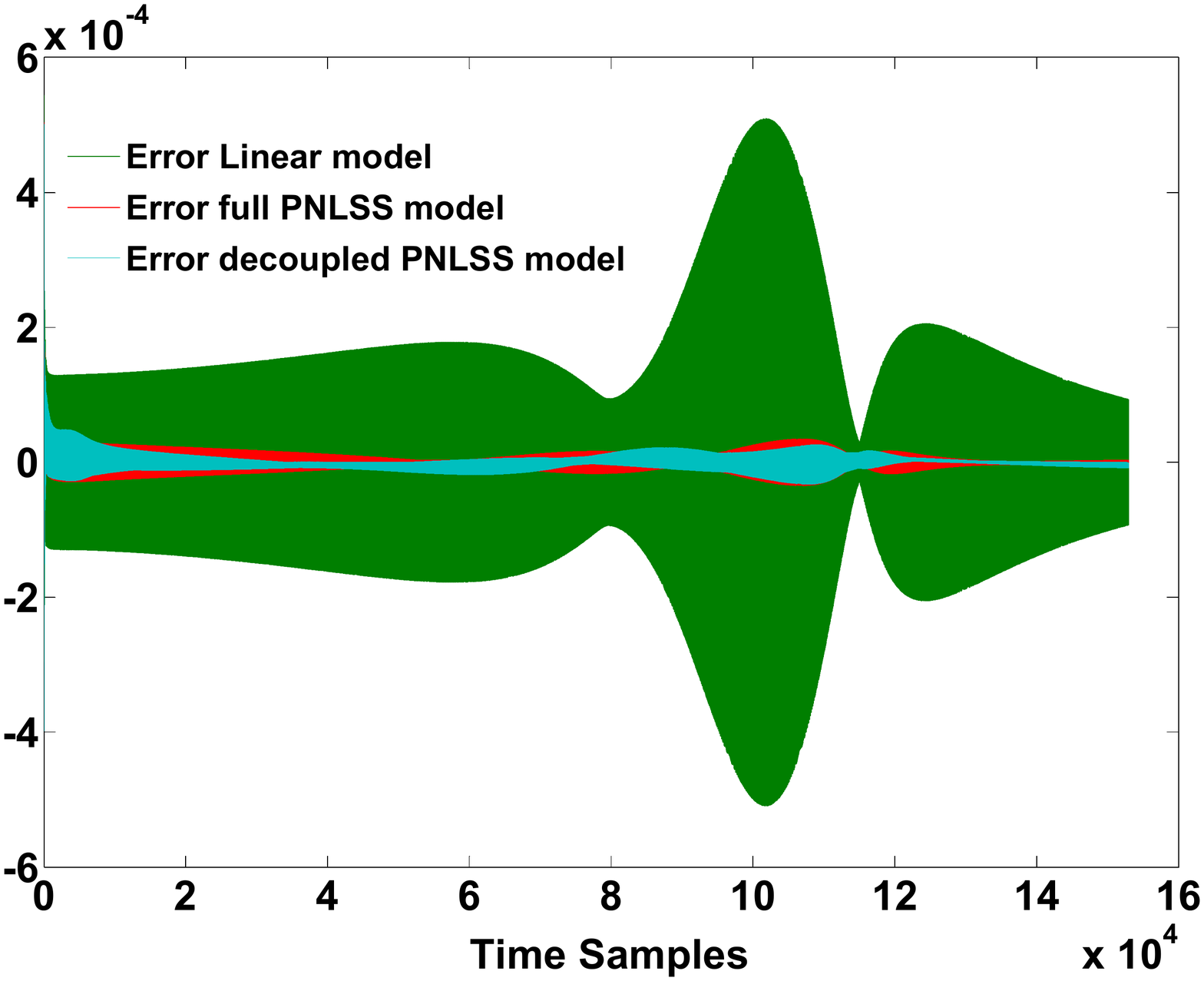}    
\caption{Time series of the output response error of the linear system (in green), PNLSS model (in red), and the decoupled model (in cyan) with 3 branches and nonlinear degree up to 10, on the swept--sine test data. The frequency band of input swept--sine is $[20 - 50]$~Hz. The natural frequency of the system is $35.59$~Hz \citep{Maarten}.
The full and decoupled PNLSS model can significantly improve on the linear model's prediction of the output to a swept sine, especially around the resonance frequency.}
\label{fig:Time}
\end{center}
\end{figure}

\begin{figure}[ht]
\begin{center}
\includegraphics[scale = .35, trim =2.5cm 3.3cm 0 1.9cm]{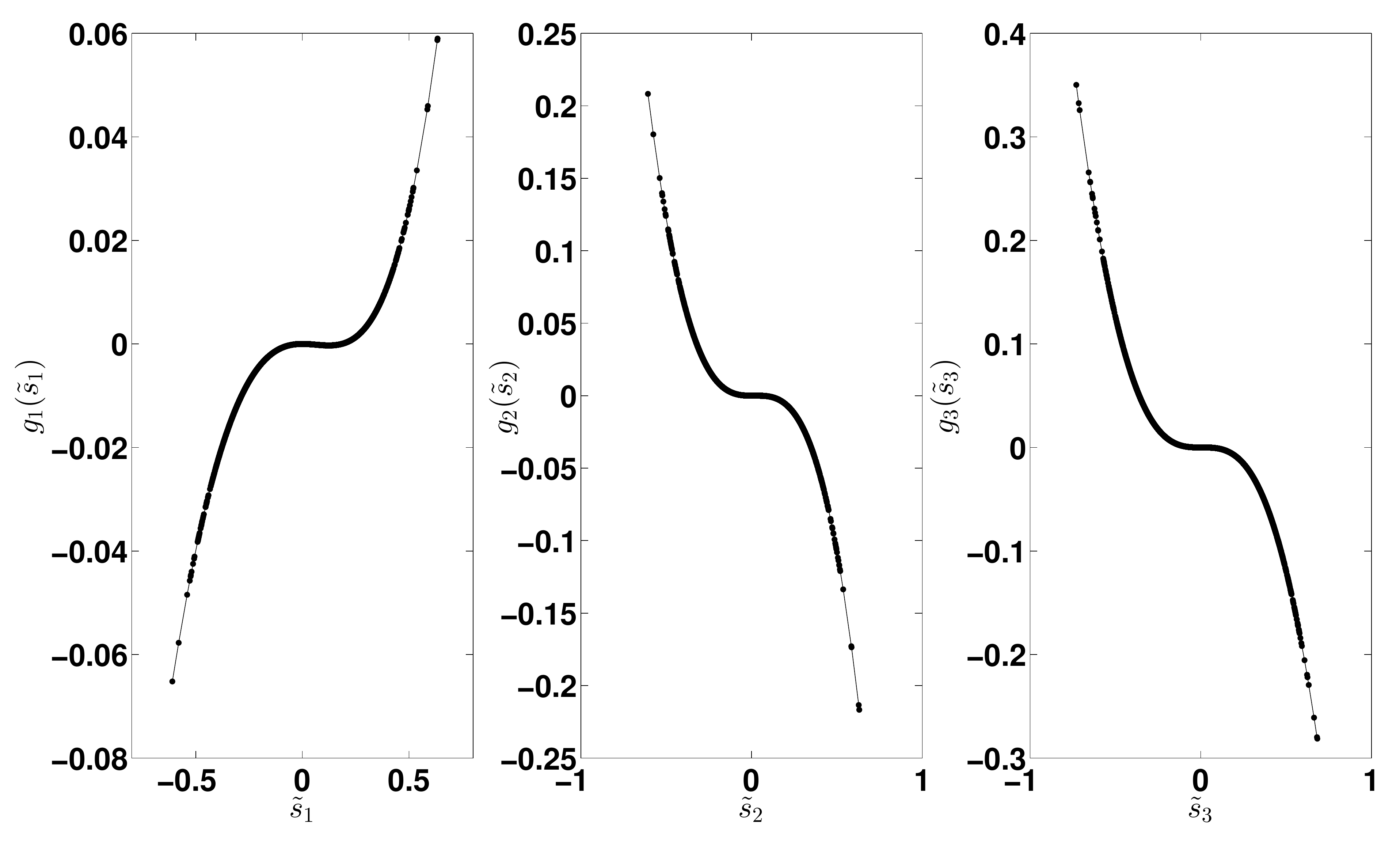}    
\caption{Three univariate polynomials in the decoupled model. The dots indicate the points where these polynomials were evaluated during the filtering on the multisine test data.
'Explosive' behavior of the polynomials is avoided here by training the model on data with a slightly larger excitation amplitude than the test data.} 
\label{fig:Br}
\end{center}
\end{figure}

Among all of these models the results of the decoupled model with the minimum amount of rms error in the test data (obtained for a model with 3 branches and $10^{th}$ order of nonlinearity) is shown in Figures \ref{fig:FRF}, \ref{fig:Time}, and \ref{fig:Br}. The output spectrum of the multisine test data is plotted in Figure \ref{fig:FRF} together with the output error spectra of the linear, full PNLSS, and decoupled PNLSS models. The PNLSS error shows a significantly higher error at higher frequencies than that of the decoupled model. The PNLSS and decoupled model have a $20$~dB lower error around the resonance frequency than the linear model. The decoupled model has even lower error than the PNLSS model in the higher frequency part of the spectrum (higher than the resonance frequency). The time series of the output error of the three models (linear, PNLSS, and decoupled model) is plotted in Figure \ref{fig:Time} for the swept-sine test data. The full and decoupled PNLSS model have a lower error than the linear model, especially around the resonance frequency, which in the swept-sine data is reached after about 10000 samples.

The decoupled model's univariate polynomials are shown in Figure \ref{fig:Br}. Although the univariate polynomials look very smooth, the polynomials from the CPD may show some outliers. In other words there would be some points far from the cluster of other points. The drawback of using polynomials is their '\textit{explosive}' behavior outside the domain where they were estimated. In future works we will consider the use of other basis functions, such as splines or neural networks.

\subsection{Some technical issues}\label{SubSec:TechIss}
The authors tried two possible alternatives in the decoupling algorithm presented in Section \ref{sec:DecouplingAlgorithm}, but the results were not satisfactory.

The first alternative is related to the first step in the decoupling algorithm. Instead of evaluating the Jacobian in random sampling points, the sampling points were drawn from the states and input of the full PNLSS model. 
Intuitively, this strategy tries to make the approximation of the multivariate polynomial with the decoupled polynomials as good as possible in (some of) the samples where the full polynomial is evaluated. But the performance of the decoupled models in this case was not satisfactory. The rms errors ranged from $-85.20$~dB to $-88.76$~dB in 5 trials, which is significantly higher than the $-94.56$~dB rms error of the full PNLSS model and the rms errors ranging from $-97.47$~dB to $-84.18$~dB of the decoupled models obtained after evaluating the Jacobian in random sampling points (see also Figure \ref{fig:RMS1}).

The second alternative is related to the observation that the decoupled 3-branch models with $10^{th}$ order nonlinearities can perform very well (the best model has $-97$~dB error on the multisine test data), but also quite poor (the worst model has about $-84$~dB error). The 3-branch models with $3^{rd}$ order nonlinearities don't have so much variability on the error (best: $-90$~dB, worst: $-87$~dB). In the second alternative, the 3-branch models with $10^{th}$ order nonlinearities were initialized from the 3-branch models with $3^{rd}$ order nonlinearities, i.e. with the higher-degree coefficients initialized as zeros. With this approach, however, the models got stuck in a local minimum with an error close to that of the initial 3-branch models with $3^{rd}$ order nonlinearities.

\section{Conclusion and future work}\label{Sec:Concl}
In this article, an algorithm was proposed to decouple the polynomial state-space model. 
We found that decoupling a full PNLSS model was able to reduce the number of model parameters, while maintaining a comparable error level.  
Several models of different orders were found that could achieve error levels comparable to (and even smaller than) those of the full PNLSS model. 
The decoupled model showed an acceptable behavior, not only on the multisine test data set (the training data set also consisted of multisine data), but also for the swept-sine class of test data.
As a future work the authors are looking for a process to select the best model out of all possible candidates. 
The physical interpretability would be non-ignorable subject to follow.
\section{Acknowledgements}
This work was supported in part by the Fund for Scientific
Research (FWO-Vlaanderen), by the Flemish Government (Methusalem), the Belgian Government through the Inter University Poles of Attraction (IAP VII) Program, and by the ERC advanced grant SNLSID, under Contract
320378.
\FloatBarrier

\section*{References}
\bibliography{elsarticle}

\begin{thebibliography}{10}
\expandafter\ifx\csname url\endcsname\relax
  \def\url#1{\texttt{#1}}\fi
\expandafter\ifx\csname urlprefix\endcsname\relax\def\urlprefix{URL }\fi
\expandafter\ifx\csname href\endcsname\relax
  \def\href#1#2{#2} \def\path#1{#1}\fi

\bibitem{bern}
J.~Oh, B.~Drincic, D.~S. Bernstein, Nonlinear feedback models of hysteresis,
  IEEE Control Systems 29~(1) (2009) 100--119.

\bibitem{bern1}
D.~Bernstein, Ivory ghost [ask the experts], IEEE Control Systems 27~(5) (2007)
  16--17.

\bibitem{hassani2014survey}
V.~Hassani, T.~Tjahjowidodo, T.~N. Do, A survey on hysteresis modeling,
  identification and control, Mechanical systems and signal processing 49~(1)
  (2014) 209--233.

\bibitem{Ikhouane11}
F.~Ikhouane, J.~Rodellar, Systems with Hysteresis Analysis, Identification and
  Control Using the Bouc-Wen Model, John Wiley \& Sons, 2007.

\bibitem{Noel}
J.-P. No\"{e}l, A.~{Fakhrizadeh Esfahani}, G.~Kerschen, J.~Schoukens, A
  nonlinear state-space approach to hysteresis identification, Mechanical
  Systems and Signal Processing 84, Part B (2017) 171--184.

\bibitem{worden2007identification}
K.~Worden, C.~Wong, U.~Parlitz, A.~Hornstein, D.~Engster, T.~Tjahjowidodo,
  F.~Al-Bender, D.~Rizos, S.~Fassois, Identification of pre-sliding and sliding
  friction dynamics: Grey box and black-box models, Mechanical Systems and
  Signal Processing 21~(1) (2007) 514--534.

\bibitem{worden2012identificationI}
K.~Worden, G.~Manson, On the identification of hysteretic systems. {Part I:
  Fitness} landscapes and evolutionary identification, Mechanical Systems and
  Signal Processing 29 (2012) 201--212.

\bibitem{worden2012identificationII}
K.~Worden, W.~Becker, On the identification of hysteretic systems. {Part II:
  Bayesian} sensitivity analysis and parameter confidence, Mechanical Systems
  and Signal Processing 29 (2012) 213--227.

\bibitem{worden2012parameterIII}
K.~Worden, J.~J. Hensman, Parameter estimation and model selection for a class
  of hysteretic systems using {Bayesian} inference, Mechanical Systems and
  Signal Processing 32 (2012) 153--169.

\bibitem{li2004improvement}
S.~Li, Y.~Suzuki, M.~Noori, Improvement of parameter estimation for non-linear
  hysteretic systems with slip by a fast {Bayesian} bootstrap filter,
  International Journal of Non-linear Mechanics 39~(9) (2004) 1435--1445.

\bibitem{billings2013nonlinear}
S.~A. Billings, Nonlinear system identification: NARMAX methods in the time,
  frequency, and spatio-temporal domains, John Wiley \& Sons, 2013.

\bibitem{schon2011system}
T.~B. Sch{\"o}n, A.~Wills, B.~Ninness, System identification of nonlinear
  state-space models, Automatica 47~(1) (2011) 39--49.

\bibitem{Johan1}
R.~Pintelon, J.~Schoukens, System Identification: a Frequency Domain Approach,
  John Wiley \& Sons, 2012.

\bibitem{Johan2}
J.~Schoukens, R.~Pintelon, Y.~Rolain, Mastering System Identification in 100
  Exercises, John Wiley \& Sons, 2012.

\bibitem{nelles2013nonlinear}
O.~Nelles, Nonlinear system identification: from classical approaches to neural
  networks and fuzzy models, Springer Science \& Business Media, 2013.

\bibitem{Paduart1}
J.~Paduart, L.~Lauwers, J.~Swevers, K.~Smolders, J.~Schoukens, R.~Pintelon,
  Identification of nonlinear systems using polynomial nonlinear state space
  models, Automatica 46 (2010) 647--656.

\bibitem{Anna}
A.~Marconato, J.~Sj\"oberg, J.~A.~K. Suykens, J.~Schoukens, Improved
  initialization for nonlinear state-space modeling, IEEE Transactions on
  Instrumentation and Measurement 63 (2014) 972--980.

\bibitem{xie2013identification}
S.~Xie, Y.~Zhang, C.~Chen, X.~Zhang, Identification of nonlinear hysteretic
  systems by artificial neural network, Mechanical Systems and Signal
  Processing 34~(1) (2013) 76--87.

\bibitem{Philippe111}
P.~Dreesen, M.~Ishteva, J.~Schoukens, Decoupling multivariate polynomials using
  first-order information and tensor decompositions, SIAM Journal on Matrix
  Analysis and Applications 36~(2) (2015) 864--879.

\bibitem{Carrol}
J.~D. Carroll, J.-J. Chang, Analysis of individual differences in
  multidimensional scaling via an n-way generalization of \text{"Eckart-Young"}
  decomposition, Psychrometrika 35 (1970) 283--319.

\bibitem{Harshman}
R.~A. Harshman, Foundations of the {PARAFAC} procedure: Models and conditions
  for an "explanatory" multi-modal factor analysis, UCLA Working Papers in
  Phonetics 16 (1970) 1--84.

\bibitem{Kolda}
T.~G. Kolda, B.~W. Bader, Tensor decompositions and applications, SIAM Review
  51, No. 3 (2009) 455--500.

\bibitem{Fakhrizadeh}
A.~{Fakhrizadeh Esfahani}, P.~Dreesen, K.~Tiels, J.-P. No\"el, J.~Schoukens,
  Polynomial state-space model decoupling for the identiﬁcation of hysteretic
  systems, in: 20th IFAC World Congress, Toulouse, France, 2017 (Accepted).

\bibitem{Bouc}
R.~Bouc, Forced vibrations of a mechanical system with hysteresis, in:
  Proceedings of the 4th Conference on Nonlinear Oscillations, Prague,
  Czechoslovakia, 1967.

\bibitem{wen1976method}
Y.-K. Wen, Method for random vibration of hysteretic systems, Journal of the
  Engineering Mechanics Division 102~(2) (1976) 249--263.

\bibitem{Maarten}
J.~P. No\"{e}l, M.~Schoukens, Hysteretic benchmark with a dynamic nonlinearity,
  in: Workshop on nonlinear system identification benchmarks, pp. 7-14,
  Brussels, Belgium, 2016.

\bibitem{Paduart3}
J.~Paduart, J.~Schoukens, R.~Pintelon, T.~Coen, Nonlinear state space modelling
  of multivariable systems, Vol.~39, Elsevier, 2006, pp. 565--569.

\bibitem{Paduart4}
J.~Paduart, Identification of nonlinear systems using polynomial nonlinear
  state space models, Ph.D. thesis, Vrije Universiteit Brussel (VUB) (2007).

\bibitem{Laurent1}
A.~{Van Mulders}, L.~Vanbeylen, Comparison of some initialisation methods for
  the identification of nonlinear state-space models, in: IEEE International
  Instrumentation and Measurement Technology Conference ({$I^2$MTC}),
  Minneapolis, MN, 2013, pp. 807--811.

\bibitem{Laurent2}
A.~{Van Mulders}, L.~Vanbeylen, J.~Schoukens, Robust optimization method for
  the identification of nonlinear state-space models, in: IEEE International
  Instrumentation and Measurement Technology Conference ({$I^2$MTC}), Graz,
  Austria., 2012, pp. 1423--1428.

\bibitem{Rik1}
R.~Pintelon, Frequency-domain subspace system identification using
  non-parametric noise models, Automatica 38 (2002) 1295--1311.

\bibitem{van2012subspace}
P.~Van~Overschee, B.~De~Moor, Subspace identification for linear systems:
  Theory--Implementation--Applications, Springer Science \& Business Media,
  2012.

\bibitem{kyprianou2001identification}
A.~Kyprianou, K.~Worden, M.~Panet, Identification of hysteretic systems using
  the differential evolution algorithm, Journal of Sound and Vibration 248~(2)
  (2001) 289--314.

\bibitem{newmark1959method}
N.~M. Newmark, A method of computation for structural dynamics, Journal of the
  Engineering Mechanics Division 85~(3) (1959) 67--94.

\bibitem{geradin2014mechanical}
M.~G{\'e}radin, D.~J. Rixen, Mechanical vibrations: theory and application to
  structural dynamics, John Wiley \& Sons, 2014.

\bibitem{jeffreys1956methods}
H.~Jeffreys, B.~Swirles, P.~M. Morse, Methods of mathematical physics, AIP,
  1956.

\bibitem{janczak2004identification}
A.~Janczak, Identification of nonlinear systems using neural networks and
  polynomial models: a block-oriented approach, Vol. 310, Springer Science \&
  Business Media, 2004.

\bibitem{TensorLab}
N.~Vervliet, O.~Debals, L.~Sorber, M.~V. Barel, L.~{De Lathauwer},
  Tensorlab~3.0, Available online. URL: \url{http://www.tensorlab.net}, March
  2016.

\end{thebibliography}
\end{document}